\documentclass[prl,aps,twocolumn,superscriptaddress]{revtex4-1}
\pdfoutput=1
\usepackage{amssymb}
\usepackage{amsmath}
\usepackage{subfigure}
\usepackage{graphicx}
\usepackage{textcomp}
\usepackage{url}
\usepackage{color}
\usepackage{cases}
\usepackage[normalem]{ulem}

\usepackage[colorlinks=true, urlcolor=blue, anchorcolor=blue, citecolor=blue,filecolor=blue,linkcolor=blue,menucolor=blue
]{hyperref}

\usepackage[titletoc,title]{appendix}

\usepackage{color}

\begin{document}
\title{Optimal paths of non-equilibrium stochastic fields: the Kardar-Parisi-Zhang interface as a test case}

\author{Alexander K. Hartmann}
\email{a.hartmann@uni-oldenburg.de}
\affiliation{Institut f\"{u}r Physik, Universit{\aa}t Oldenburg - 26111 Oldenburg, Germany}
\author{Baruch Meerson}
\email{meerson@mail.huji.ac.il}
\affiliation{Racah Institute of Physics, Hebrew University of
Jerusalem, Jerusalem 91904, Israel}
\author{Pavel Sasorov}
\email{pavel.sasorov@gmail.com}
\affiliation{Institute of Physics CAS -- ELI Beamlines, 182 21 Prague, Czech Republic}
\affiliation{Keldysh Institute of Applied Mathematics, Moscow 125047, Russia}


\begin{abstract}

Atypically large fluctuations in macroscopic non-equilibrium systems continue to attract interest. Their probability can often be determined by the optimal fluctuation method (OFM).
The OFM  brings about a conditional variational problem, the solution of which
describes the ``optimal path" of the system which dominates the contribution of different stochastic paths to the desired statistics. The OFM proved efficient in evaluating the probabilities of rare events in a host of systems. However,  theoretically predicted optimal paths were observed in stochastic simulations only in diffusive lattice gases, where the predicted optimal density patterns are either stationary, or travel with constant speed.  Here we focus on the one-point height distribution of the paradigmatic Kardar-Parisi-Zhang interface. Here the optimal paths, corresponding to the distribution tails  at short times, are intrinsically non-stationary and can be predicted analytically. Using the mapping to the directed polymer in a random potential at high temperature, we obtain ``snapshots" of the optimal paths in Monte-Carlo simulations which probe the tails with an importance sampling algorithm. For each tail we observe a very narrow ``tube" of height profiles around a single optimal path which agrees with the analytical prediction. The agreement holds even at long times, supporting earlier assertions of the validity of the OFM in the tails well beyond the weak-noise limit.

\end{abstract}

\maketitle


\nopagebreak

Non-equilibrium behaviors of stochastic macroscopic systems are ubiquitous in nature.  One interesting group of questions about such systems concern atypically large fluctuations, which manifest themselves as tails of probability distributions of different fluctuating quantities. Although a universal description of the tails is not likely to emerge, a great progress has been achieved in the last two decades in particular systems. Much of the progress has been due to a method which emerged in different areas of physics under different names: the optimal fluctuation  method (OFM), the instanton method, the weak noise theory, the macroscopic fluctuation theory, \textit{etc}. A key ingredient of this method is a saddle-point evaluation of the path integral of the stochastic process in question, conditioned on the specified large deviation.  This evaluation, which employs a model-dependent small parameter (colloquially called ``weak noise"), brings about a conditional variational problem. Its solution -- a deterministic field, evolving in time  -- describes the ``optimal path": the most probable history of the system which dominates the contribution to the desired statistics. The OFM has been proved efficient in a host of non-equilibrium problems: turbulence and turbulent
transport \cite{Falkovich1996,GM1996,FalkovichRMP,Chernykh,Grafke,Bouchet}, stochastic surface growth and related systems \cite{Mikhailov1991,Fogedby1998,KK2007,KK2008,KK2009,FogedbyRen,MV2016,MKV,KMSparabola,Janas2016, SMS2017,
MeersonSchmidt2017, SMS2018,  SKM2018, MSV_3d, SmithMeerson2018, MV2018,Asida2019,SMV2019}, diffusive transport in the absence \cite{Jona2006,Jona2007,bertini2015,BD2004,Bertini2005,BD2005,Sukhorukov2004,Tailleur1,Appert,BDL2008,HG2009,
DG2009b,HG2011,LGW,Bunin,KM2012,Gorissen2012,void,Akkermans,MSKMP,Espirages2013,Hurtado2013review,MSSSEP,Sadhu,VMS2014,MR,MVK,AMV,ZM,Espirages2016,Baek,AM2017,Baek2,
ZM,Baek,AM2017,Baek2,narrow,AKM2019,DerridaSadhu2019}
and in the presence \cite{Elgart,Basile2004,Bodieneau2010,MS2011a,MSKaplan,MS2011b,MVS,Hurtado2013,M2015,Lasanta} of particle reactions or energy dissipation, \textit{etc}.  However, we are aware only of one class of systems -- diffusive lattice gases -- where theoretically predicted optimal paths were actually observed in stochastic simulations \cite{HG2009,HG2011,Gorissen2012,Hurtado2013review,ZM,Espirages2016}. Furthermore, in all these systems the optimal density patterns are either stationary, or travel with constant speed.

Here we report (to our knowledge, for the first time) the observation of intrinsically non-stationary optimal paths. We consider the paradigmatic  Kardar-Parisi-Zhang (KPZ) equation in 1+1 dimension \cite{KPZ}:
\begin{equation}
\partial_{t}h=
\nu\partial_{x}^{2}h+(\lambda/2)\left(\partial_{x}h\right)^{2}+\sqrt{D}\,\xi(x,t),\label{KPZ1}
\end{equation}
where $\nu$ is the diffusivity, and we can set  the non-linearity coefficient $\lambda>0$ without loss of generality. Equation (\ref{KPZ1}) describes the evolution of the height $h(x,t)$ of an interface driven by the Gaussian noise $\sqrt{D}\xi(x,t)$ which has zero mean and is delta-correlated: $\langle\xi(x_{1},t_{1})\xi(x_{2},t_{2})\rangle=\delta(x_{1}-x_{2})\delta(t_{1}-t_{2})$.
The KPZ equation has been extensively studied \cite{KPZ,HHZ,Barabasi,Krug,S2016}. At long times,
the interface width was found to grow as $t^{1/3}$, while the lateral correlation
length grows as $t^{2/3}$. The exponents $1/3$ and $2/3$ are the
hallmark of a whole universality class of the 1+1 dimensional non-equilibrium
growth \cite{HHZ,Barabasi,Krug,Corwin,QS,S2016,Takeuchi2017}.

In the last decade the studies of the KPZ equation and related systems have seen spectacular developments \cite{Corwin,QS,S2016,Takeuchi2017}. They shifted from the interface roughness and correlation function to more detailed characteristics such as the full probability distribution of the interface height at a point, $\mathcal{P}\left(H,t\right)$.  Remarkably, exact representations were derived for $\mathcal{P}\left(H,t\right)$ for the KPZ equation on the line $|x|<\infty$ for several basic initial conditions: ``droplet" \cite{SS2010,Calabrese2010,Dotsenko2010,Amir2011}, flat \cite{flat}, stationary \cite{stationary1,stationary2}, and some of their combinations.  At long times and for typical height fluctuations, $\mathcal{P}\left(H,t\right)$ converges to the Tracy-Widom (TW) distribution for the Gaussian unitary ensemble (GUE) of random matrices
\cite{TWGUE} for the droplet, to the TW distribution for the Gaussian orthogonal ensemble
(GOE) \cite{TWGOE} for the flat interface, and to the Baik-Rains  (BR) distribution \cite{BR} for the stationary
interface. These theoretical predictions were confirmed in ingenious experiments with liquid-crystal turbulent fronts \cite{experiment}, see also Ref. \cite{interpretation}. Atypically large fluctuations -- the tails of $\mathcal{P}\left(H,t\right)$ -- behave differently and are unrelated to the TW and BR distributions. These tails were determined, for the three basic initial conditions and for several other initial conditions, by the OFM \cite{KK2007,KK2008,KK2009,MKV,KMSparabola,Janas2016,MeersonSchmidt2017,SKM2018}. The OFM predictions
were verified in all cases where the corresponding tails were also extracted from exact representations.

The most thoroughly studied is the droplet case, for which the OFM predicted the following tails \cite{KMSparabola}:
\begin{numcases}
{-\ln \mathcal{P}(H,t) \simeq}
\frac{8\sqrt{2}\, \nu }{3D
\lambda^{1/2}}\,\frac{H^{3/2}}{t^{1/2}}, & $H\to +\infty$,
\label{hightail}\\
\frac{4\sqrt{2\lambda}}{15 \pi
D}\,\frac{|H|^{5/2}}{t^{1/2}}, &$H\to - \infty$.
\label{lowtail}
\end{numcases}
The higher tail (\ref{hightail}) coincides with the corresponding tail of the GUE TW distribution, while the lower tail (\ref{lowtail}) is different from the GUE TW tail (the latter scales as $|H|^3/t$). Using the exact representations, it was  shown that the tails (\ref{hightail}) and  (\ref{lowtail}) are observed both at short times \cite{DMRS2016} and at long times \cite{DMS2016,SMP,Corwinetal,CorwinGhosal2018a,KLD2018,Tsai,CorwinGhosal2018b}.
Furthermore, these tails have been recently verified in rare-event stochastic simulations \cite{Hartmann2018} which employed a standard mapping to (a discrete version of) the directed polymer in a random potential at high temperature \cite{Huse1985,KPZ,HHZ,S2016,Krug,CLDR2010}. The problem of the KPZ one-point height statistics in 1+1 dimension for the droplet initial condition can therefore serve as a good test case for the studies of \emph{optimal paths} in spatially explicit and non-stationary stochastic systems conditioned on large deviations. Here we use the directed polymer mapping and the importance sampling algorithm \cite{Hartmann2018} to obtain ``snapshots" of the optimal paths. As in the previous work  \cite{Hartmann2018}, our simulations probe the distribution tails with reaching the probability densities as small as $10^{-1000}$.  For each of the two tails we observe a very narrow tube of height profiles around a single optimal path which agrees with an analytical prediction. The agreement holds even at relatively large times, supporting earlier assertions of the validity of the OFM in the tails well beyond the weak-noise limit.

\textit{1. Optimal path: theoretical predictions for $h(x,t/2)$.}  Let us recall the OFM formulation of the problem of one-point KPZ height statistics \cite{KK2007,MKV,KMSparabola}.  For typical fluctuations, the OFM relies on the smallness of the dimensionless parameter $\epsilon =(t/t_{\text{NL}})^{1/2}\ll 1$, where $t_{\text{NL}} =\nu^5/(D^2 \lambda^4)$  is the characteristic nonlinear time of the KPZ equation. A saddle-point evaluation of the
path integral, corresponding to Eq.~(\ref{KPZ1}), leads to a variational problem for the action. The resulting Euler-Lagrange equations can be recast in the Hamiltonian form. In the rescaled variables $\tau/t \to \tau$,  $x\sqrt{\nu t}\to x$, and $\lambda h/\nu \to h$, the Hamilton's equations are
\begin{eqnarray}
  \partial_{\tau} h &=&  \partial_{x}^2 h +(1/2) \left(\partial_x h\right)^2+\rho ,  \label{eqh}\\
  \partial_{\tau}\rho &=& - \partial_{x}^2 \rho + \partial_x\left(\rho \partial_x h\right) ,\label{eqrho}
\end{eqnarray}
where $\rho(x,\tau)$ is the ``momentum" density field, canonically conjugate to $h(x,\tau)$; it describes the optimal realization of the noise $\xi(x,t)$.  In Ref. \cite{KMSparabola} a parabolic initial condition $h(x,0)=-x^2/L$ was considered. The droplet case follows from these results in the limit of $L\to 0$.  The condition  $h(x=0,\tau=t)=H$ translates into
$\rho(x=0,\tau=t)=\Lambda \,\delta(x)$; the Lagrange multiplier $\Lambda$ is ultimately expressed through $H$. Once the OFM problem is solved, one can calculate the (rescaled) action $s  =s(H) = (1/2) \int_0^t d\tau \int  dx\,\rho^2 (x,\tau)$
and obtain, in the original variables,
\begin{equation}\label{action}
-\ln \mathcal{P}(H,t) \simeq \frac{\nu^{5/2}}{D\lambda^2\sqrt{t}}\,\,s\left(\frac{\lambda H}{\nu}\right).
\end{equation}
As the noise magnitude $D$ drops out from the ``classical mechanics" formulation, the optimal path -- the solution of the OFM problem for $h(x,\tau)$ -- is independent of $D$.

\begin{figure} [ht]
\includegraphics[width=0.30\textwidth,clip=]{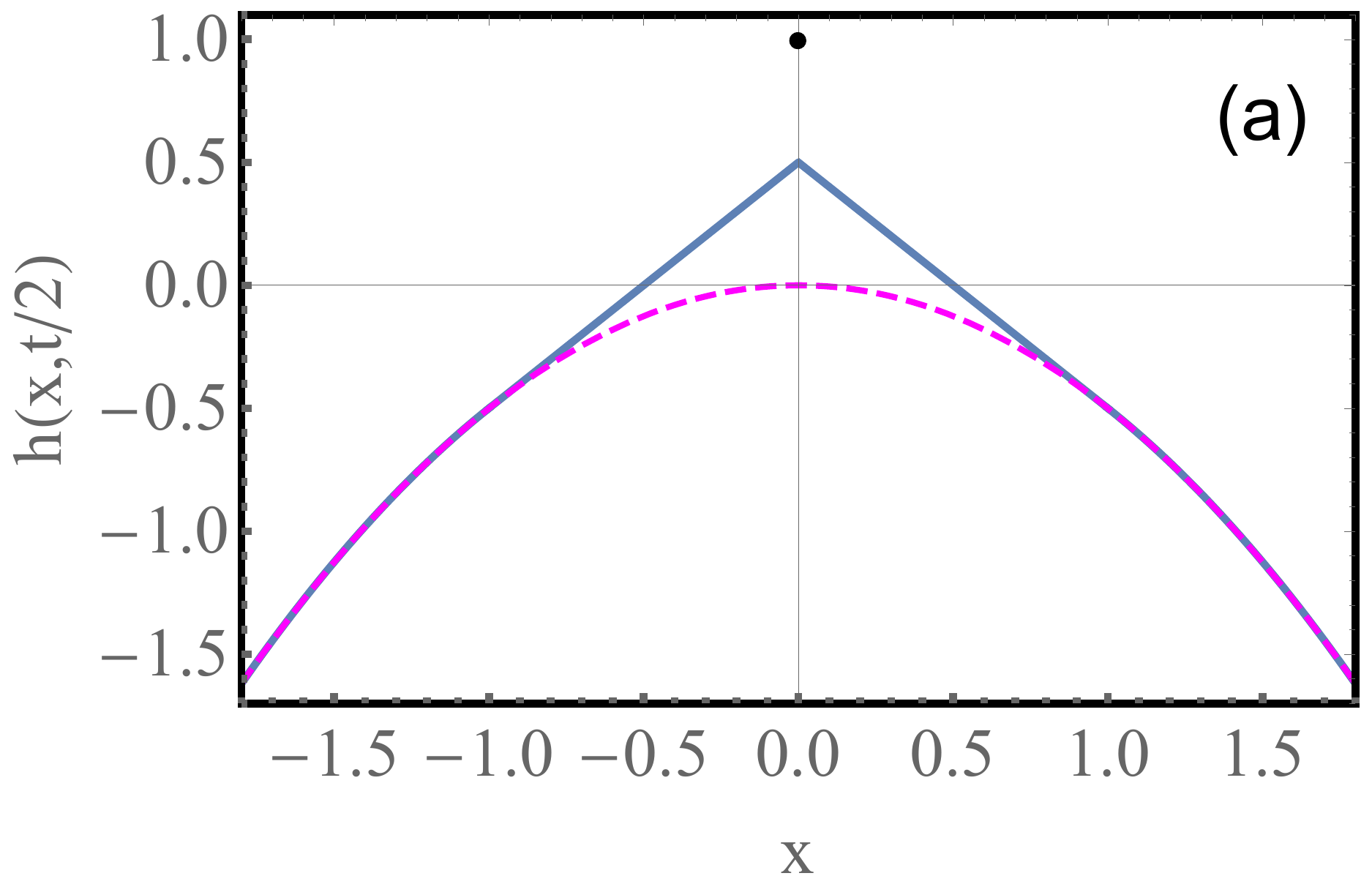}
\includegraphics[width=0.30\textwidth,clip=]{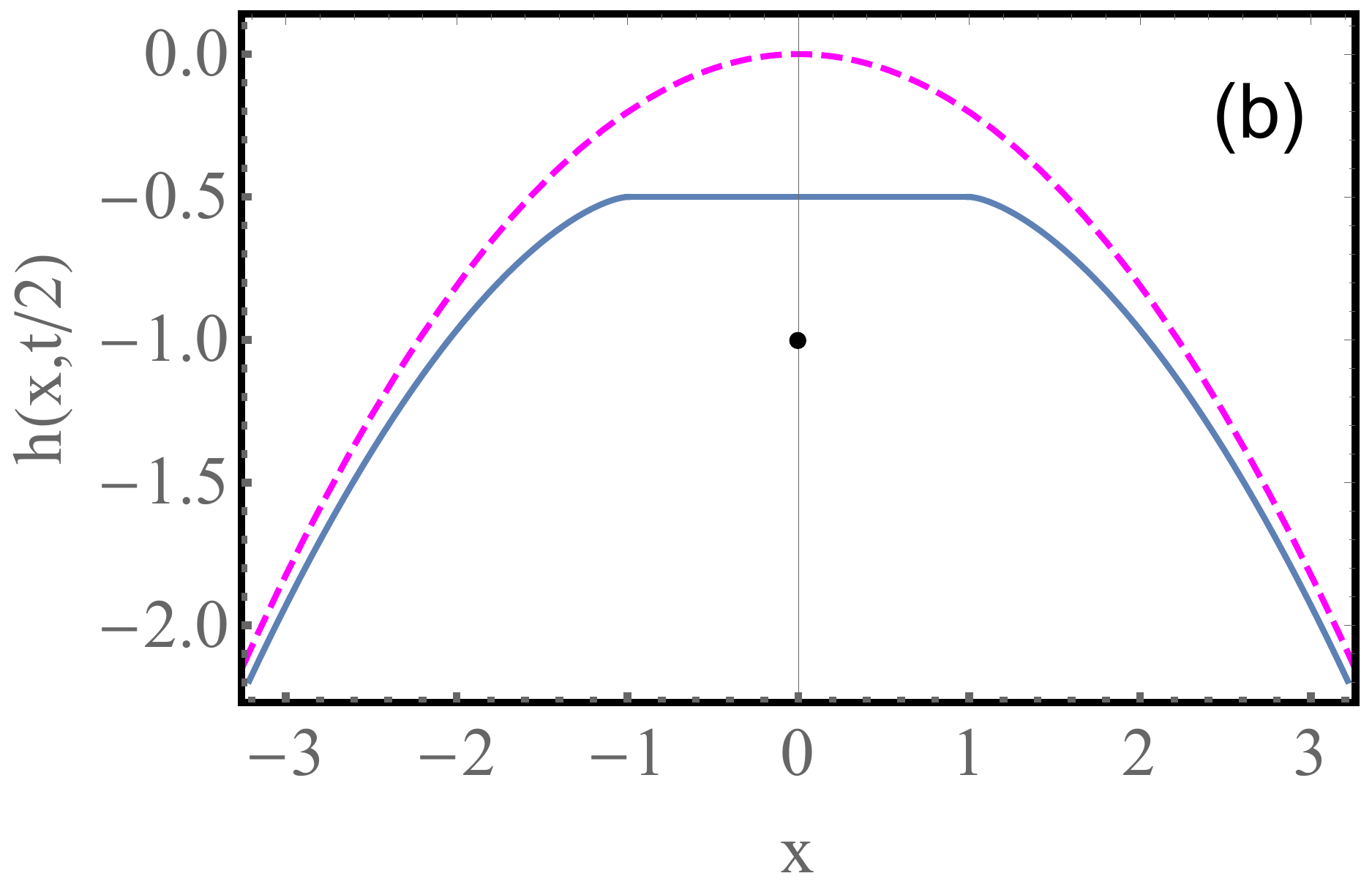}
\caption{Analytical predictions of optimal paths at $t=1/2$ conditioned on reaching  a
large positive (a) and negative (b) height $H$  at $x=0$ at time $t$ (denoted by the fat point). (a):
prediction of Eqs.~(\ref{hin2}) and~(\ref{hout2}); $x$ is rescaled by $\sqrt{\lambda H t/2}$. 
(b): prediction of Eqs.~(\ref{hin1}) and~(\ref{hout1}); $x$ is rescaled by $\sqrt{2 \lambda |H| t}/\pi$. 
In both panels $h$ is rescaled by $|H|$. The dashed lines, $-x^2/2$ (a) and  $-2 x^2/\pi^2$ (b),
show in the corresponding units the zero-noise KPZ interface at time $t/2$.}
\label{hightailsnap}
\end{figure}

As of present, the exact optimal path is unknown, but asymptotic solutions for very large positive and negative $H$ [which determine the tails (\ref{hightail}) and (\ref{lowtail})] are available \cite{KMSparabola}.  Here are ``snapshots" of these asymptotic solutions at $\tau=t/2$. For very large positive $H$  
\begin{numcases}
{\!\!h\left(x,\frac{t}{2}\right) \simeq}
\frac{H}{2}-\sqrt{\frac{2 H}{\lambda t}}\,|x|, & $|x|\leq \sqrt{\lambda
H t/2}$,
\label{hin2}\\
-x^2/\lambda t
, &$|x|>\sqrt{\lambda H t/2} $, \label{hout2}
\end{numcases}
see Fig. \ref{hightailsnap} a. In the leading order  in $\lambda H/\nu \gg 1$, $h(x,\tau)$ does not depend on the diffusivity $\nu$. Equation~(\ref{hin2}) describes a snapshot of two outgoing ``ramps" of $h(x,\tau)$, which correspond to a stationary ``antishock" of the  interface slope $v(x,\tau)=\partial_x h(x,\tau)$ at $x=0$, sustained by a very narrow stationary pulse of $\rho$ \cite{Fogedby1998,KK2007,MKV,KMSparabola}. The solution (\ref{hout2}) describes the noiseless ($\xi=0$) evolution of the KPZ equation (\ref{KPZ1}) starting from the droplet initial condition. The solutions (\ref{hin2}) and (\ref{hout2}) are continuous with their first derivatives at the moving boundaries between them \cite{KMSparabola}. At $\tau=t/2$ these boundaries are at $|x|= \sqrt{\lambda H t/2}$.
If one accounts for the diffusion, the corner singularity of $h(x,t/2)$ at $x=0$  is smoothed \cite{MKV}:
\begin{equation}\label{hin2tip}
\!\!h\!\left(\!x,\frac{t}{2}\right) \!\simeq\!
\frac{H}{2}-\frac{2\nu}{\lambda}\ln \cosh \!\left(\!\sqrt{\frac{\lambda H}{2\nu^2 t}}\,x\!\right), |x|\leq \sqrt{\frac{\lambda
H t}{2}},
\end{equation}

For very large negative $H$ the optimal path is quite different. Using the results of Ref. \cite{KMSparabola}, we obtain
\begin{numcases}
{\!\!h\left(x,\frac{t}{2}\right) \simeq}
\frac{H}{2}, & \!\!\!\!\!\!$|x|\leq \frac{\sqrt{2\lambda |H|t}}{\pi}$,
\label{hin1}\\
\frac{H}{2}\,\mathfrak{h}\left(\frac{\pi\,|x|}{\sqrt{2\lambda
|H|t}}\right), &\!\!\!\!\!\!$|x|\geq \frac{\sqrt{2\lambda |H|t}}{\pi}$,
\label{hout1}
\end{numcases}
where the function $\mathfrak{h}(z)$ is defined parametrically,
$z(u)=1+u \arctan u, \mathfrak{h}(u) = 1+\frac{2}{\pi}\left[u+(u^2-1) \arctan u\right]$,
$0\leq u<\infty$.
Here too the optimal path is independent of $\nu$ in the leading order. Noticeable is an extended flat region of $h(x,t/2)$, see Fig. \ref{hightailsnap} b. As one can see, the optimal paths for large positive and negative $H$ are strikingly different.

\begin{figure} [ht]
\includegraphics[width=0.21\textwidth,clip=]{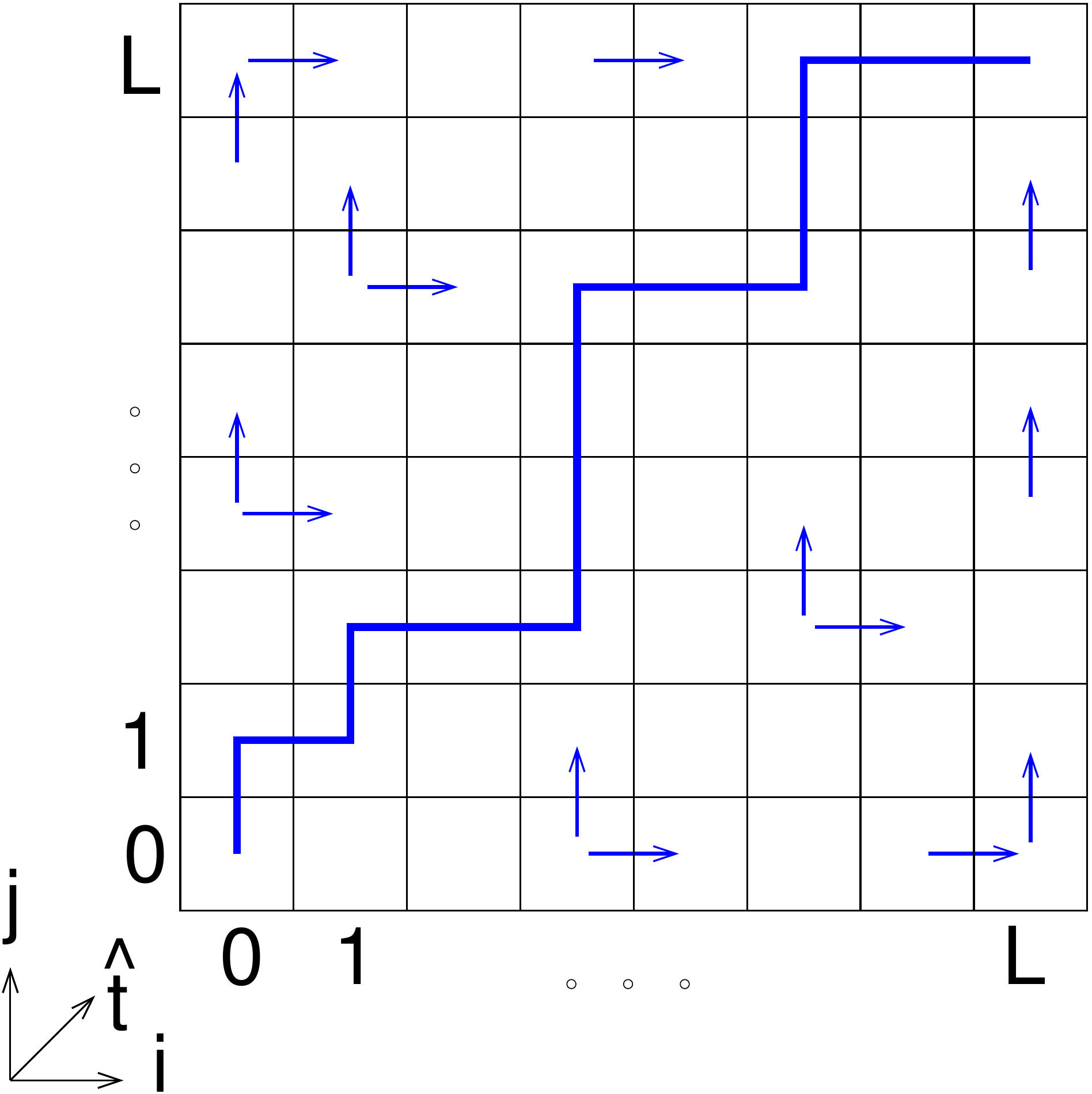}
\caption{A realization of the directed polymer $(0,0)\to(L,L)$.
 The arrows indicate samples for possible
directions of the polymer for some lattice sites.}
\label{directedpolymer}
\end{figure}

\textit{2. Directed polymer mapping.}
Now let us recall the discrete version of the mapping
between the KPZ height $h(x,t)$ and the free energy of a directed polymer in a
two-dimensional random potential  at high temperature $T$ \cite{Calabrese2010}. Consider all directed polymers which start at the point $(0,0)$ and end at the point $(L,L)$ of a square lattice indexed by integers $(i,j)$ which run from $0$ to $L$, as shown in Fig. \ref{directedpolymer}.  The value of the potential $V$ at each lattice point is normally distributed with zero mean and unit variance. Introduce the new variables $\hat{\tau}=i+j$ and $\hat{x} = (i-j)/2$.  The partition function $Z(i,j)$  of a given random configuration of the potential, $Z(i,j)=\hat{Z}[(i-j)/2,i+j]$, with the end point $(i,j)$,
obeys the exact recursive equation \cite{Calabrese2010}
\begin{equation}\label{A004}
\!\hat{Z}\left(\hat{x},\hat{\tau}+1\right)\!=\! \left[\hat{Z}\left(\hat{x}-\frac{1}{2},\hat{\tau}\right)+
\hat{Z}\left(\hat{x}+\frac{1}{2},\hat{\tau}\right)\right]\! e^{-\frac{V\left(\hat{x},\hat{\tau}+1\right)}{T}},
\end{equation}
where $\left\langle V\left(\hat{x},\hat{\tau}\right)V\left(\hat{x}^\prime,\hat{\tau}^\prime\right)\right\rangle
=\delta_{\hat{x}\hat{x}^\prime}\delta_{\hat{\tau} \hat{\tau}^\prime}$, and $\delta_{\hat{x}\hat{x}^\prime}$ and $\delta_{\hat{\tau} \hat{\tau}^\prime}$ are Kronecker deltas.
Let $Z^*\left(\hat{x},\hat{\tau}\right) = 2^{-\hat{\tau}} \hat{Z}\left(\hat{x},\hat{\tau}\right)$. The mapping to the continuous Eq.~(\ref{KPZ1}) is achieved via a truncated  Taylor expansion of all the discrete quantities and $\exp(-V/T)$ in Eq.~(\ref{A004}). For example,
\begin{equation}\label{taylor}
Z^*(\hat{x}-\frac{1}{2},\hat{\tau})=Z^*(\hat{x},\hat{\tau})
-\frac{1}{2}\partial_{\hat{x}}Z^*(\hat{x},\hat{\tau})
+\frac{1}{8}\partial_{\hat{x}}^2Z^*(\hat{x},\hat{\tau})+\dots .
\end{equation}
To justify these expansions, we use strong inequalities $L\gg 1$ and $T\gg 1$. This procedure approximates the discrete Eq.~(\ref{A004}) by the stochastic heat equation
\begin{equation}\label{A010}
\partial_{\tau} Z^*=\frac{1}{8}\partial_x^2Z^*-\frac{\xi}{T}Z^* ,
\end{equation}
where we dropped the hats of $\hat{x}$ and $\hat{\tau}$. The (now continuous) noise term $\xi$ is delta-correlated in $x$ and $\tau$. The initial condition is $Z^*(x,0) = \delta(x)$. The desired mapping is given by the relations
$h(x,\tau) = \ln [Z^*(x,\tau)/\langle Z^*(0,\tau) \rangle], \lambda=1/4, \nu =1/8$, $D=T^{-2}$ and $t= 2L$ \cite{Calabrese2010}.  The characteristic nonlinear time of the KPZ equation, $t_{\text{NL}} =\nu^5/(D^2 \lambda^4)$, becomes $T^4/2^7$.

\textit{3. Importance sampling algorithm.}
The rare-event simulation method used here is based on the idea to bias
the underlying distribution towards the outcomes of interest. It was
introduced
in computer science under the names \emph{importance sampling} or \emph{variance reduction}
\cite{hammersley1956} and has been frequently used in physics, in particular when
studying explicitly time-dependent processes with transition-path
sampling \cite{dellago1998}, cloning approaches \cite{giardina2006}
or density-matrix renormalization group \cite{gorissen2009}.
Our approach
is more straightforward and
somewhat more general, as it applies to a very broad range
of stochastic systems,
both in and out of equilibrium \cite{work_ising2014}.

To obtain realizations of the random field $V$ which correspond to extreme values of $H=H(V)$,
we do not sample the random field according to its natural Gaussian product weight $G(V)$,
but according to the modified weight $P(V)\sim G(V)\exp(-H/\Theta)$. $\Theta$ is an
auxiliary ``temperature'' parameter, which allows us to shift the peak of distributions
to smaller than ($\Theta>0$ close to zero) or larger than ($\Theta<0$ close to zero)
typical values. The random realizations cannot be generated directy according to
the weight $P(V)$. Instead we use a standard Markov-chain approach with the
Metropolis-Hastings algorithm \cite{newman1999},
where the configurations of the Markov chain are
the random realizations. By sampling for different values of $\Theta$ and combining
and normalizing the results, $H$ can be sampled over a larger range of the support
\cite{Hartmann2018}, down to probabilities like $10^{-1000}$.
For details of the approach, see Refs. \cite{align2002,work_ising2014,Hartmann2018}.
We extend the approach by storing regularly
during the simulation (after equilibration)
configurations sampled at various
values of $\Theta$. We also extend the simulations to even
smaller values of $H$.
For configurations exhibiting corresponding values of  $H$ of
interest we then can analyze $h(x,t/2)$.

\begin{figure} [ht]
\includegraphics[width=0.30\textwidth,clip=]{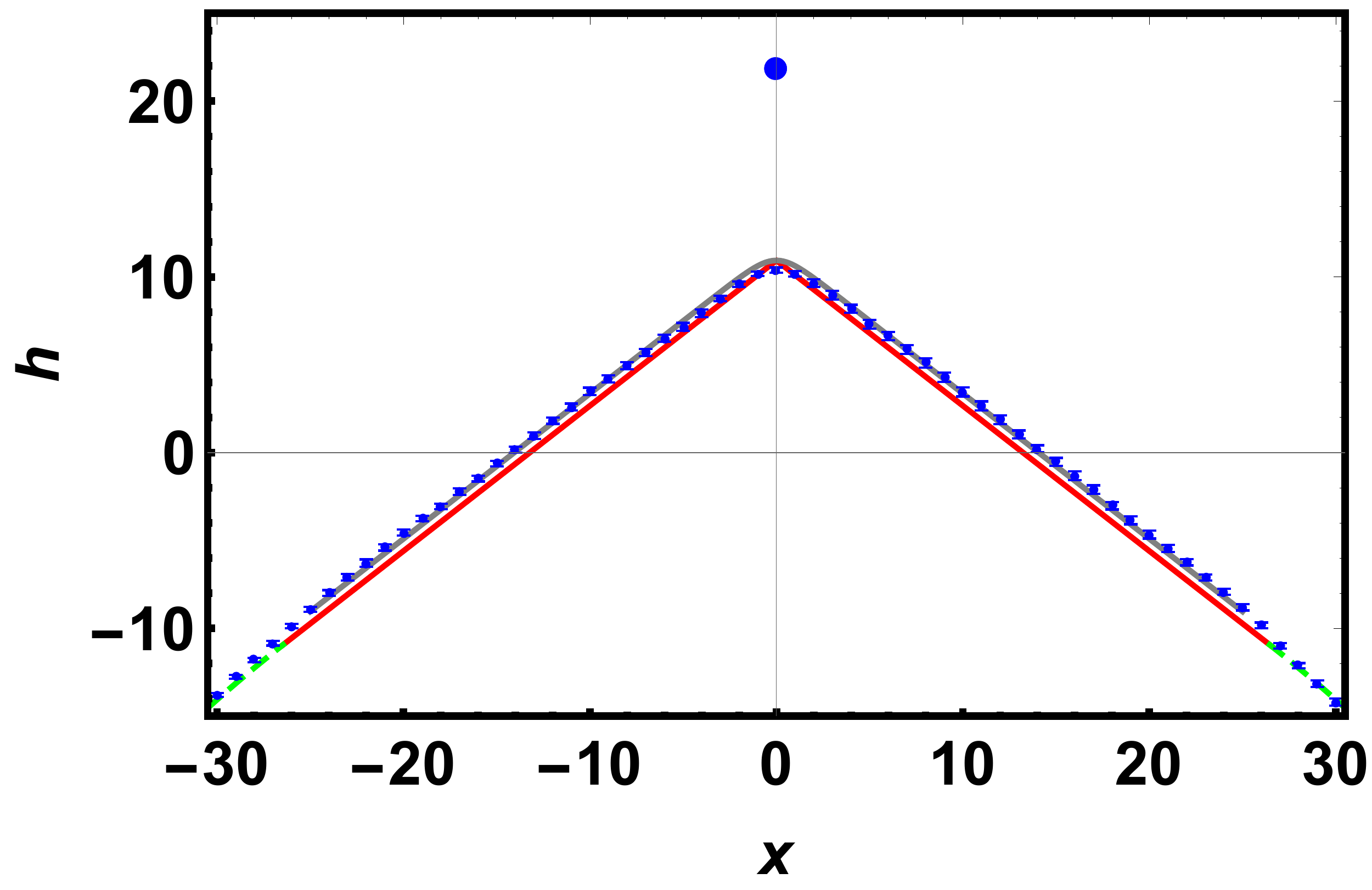}
\caption{The optimal path at $\tau = t/2$, conditioned on a large positive height $H$ (close to $21.9$, the fat point) at $\tau =t$  for $T^4=2^{13}$ and $L=2^7$. The symbols with error bars show the results of 10 realizations. The lower solid line is the zero-diffusion solution (\ref{hin2}). The higher solid line is the finite-diffusion solution (\ref{hin2tip}). Dashed line: the parabola (\ref{hout2}).}
\label{hightailpath}
\end{figure}

\textit{4. Simulation results and comparison with theory.} In our first series of simulations we chose $L=2^7$ and $T^4=2^{13}$. Although $\epsilon = (t/t_{\text{NL}})^{1/2}=2$ is not small, the simulations of Ref. \cite{Hartmann2018} confirmed the validity of the OFM for the whole distribution $\mathcal{P}(H,t)$. One can expect, therefore, that the OFM predictions of the optimal paths will be accurate. This is indeed what our simulations show.
Figure \ref{hightailpath} is a result of processing of 10 simulated configurations at $\tau = t/2$, conditioned on reaching a large positive $H$ (close to 21.9) at $\tau =t$. They appear in the natural ensemble with a probability near $10^{-200}$
\cite{Hartmann2018}. Instead of showing the 10 actual profiles, we only presented the profile averaged over the realizations and the error bars. As one can see, the error bars are strikingly small, implying a very narrow tube of stochastic trajectories around the optimal path.  Furthermore, this optimal path agrees very well with the leading-order (zero-diffusion) analytical prediction (\ref{hin2}) and (\ref{hout2}), and remarkably well with the finite-diffusion expression (\ref{hin2tip}).

\begin{figure} [ht]
\includegraphics[width=0.30\textwidth,clip=]{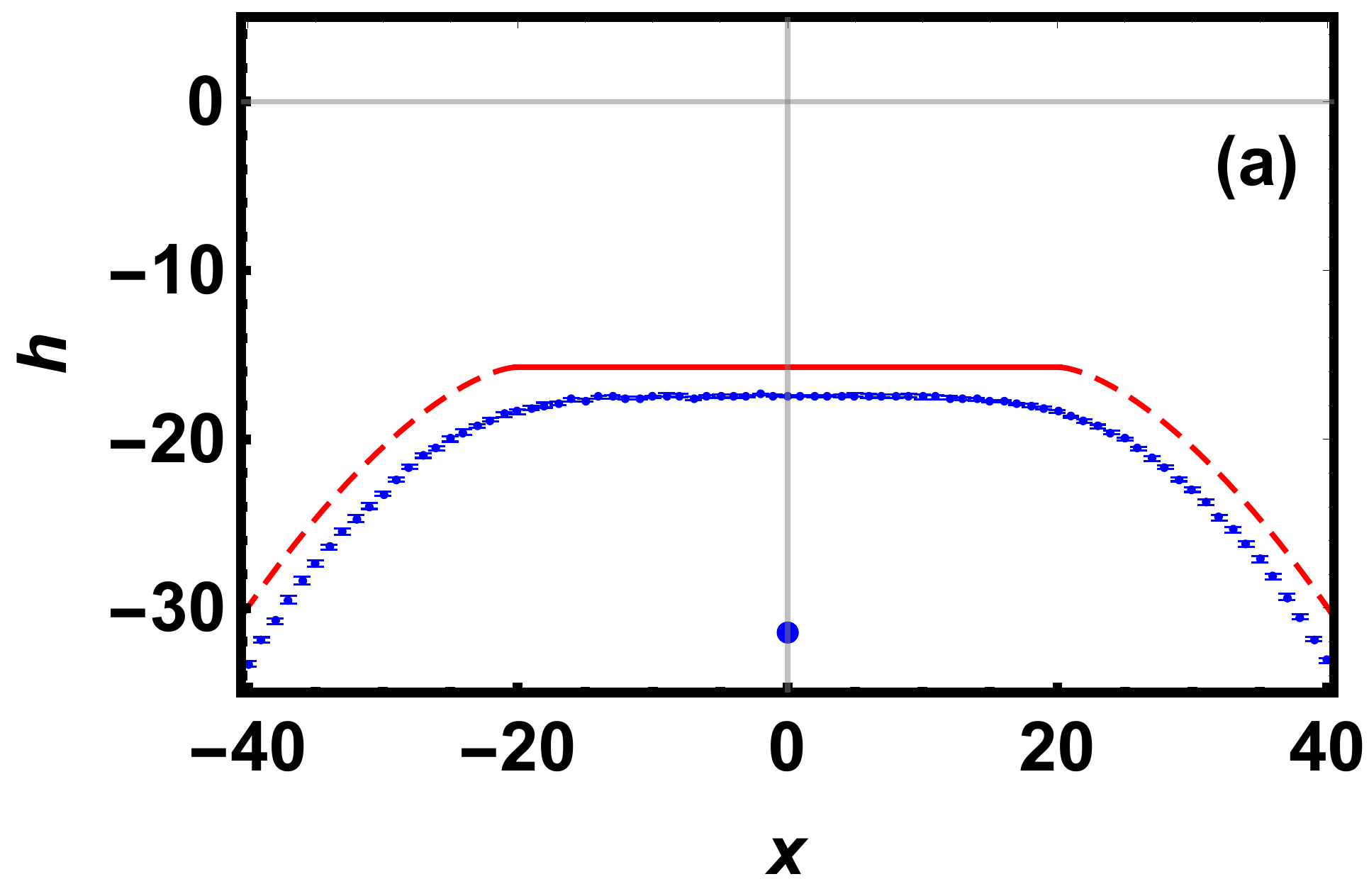}
\includegraphics[width=0.30\textwidth,clip=]{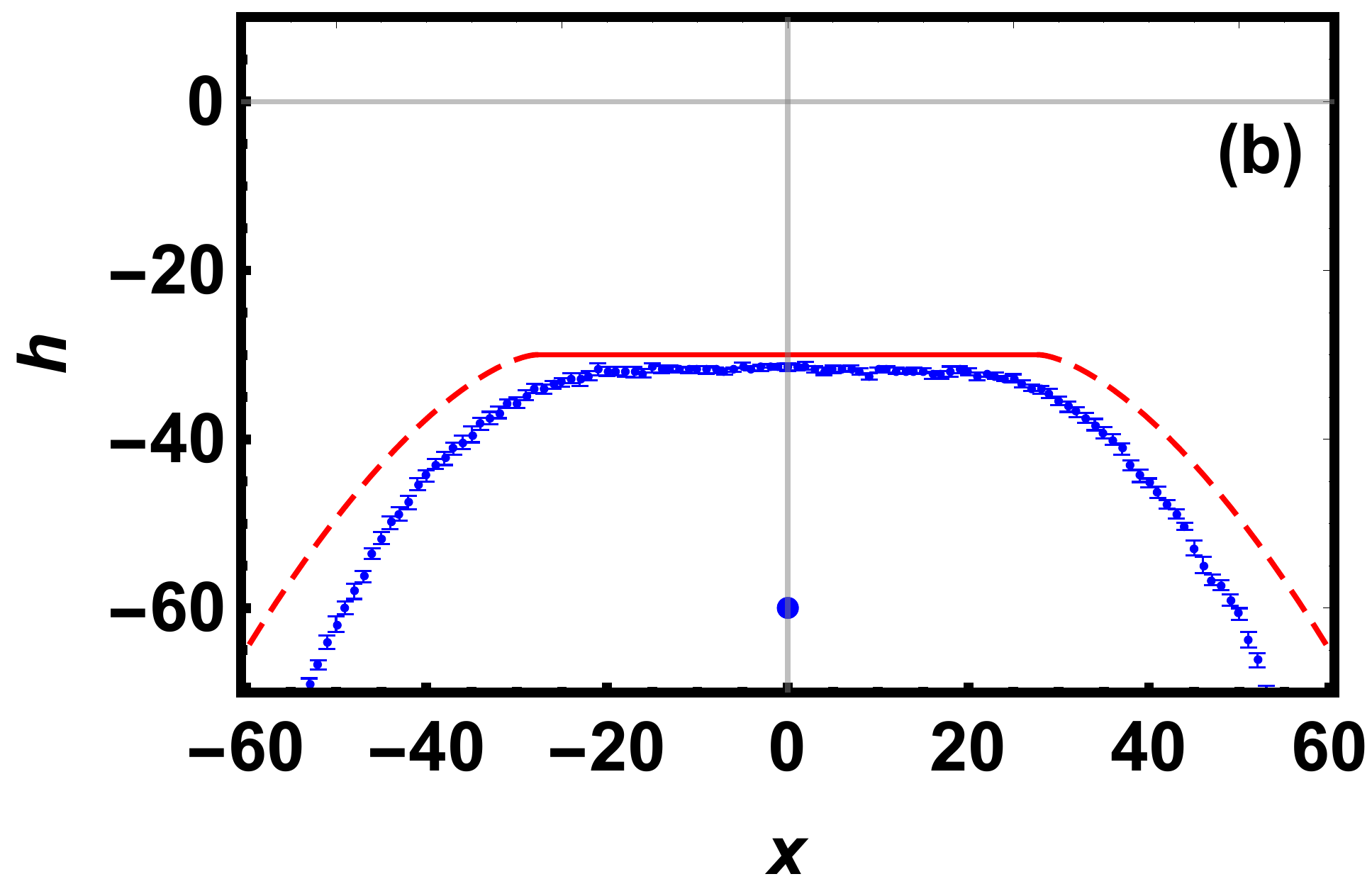}
\caption{The optimal path at $\tau = t/2$, conditioned on a large negative height $H$ (the fat point) at $\tau =t$ and $L=2^7$. The symbols with error bars show the results of 10 realizations. The solid line is the analytically predicted plateau (\ref{hin1}) and (\ref{hout1}).  The dashed line is Eq.~(\ref{hout1}). (a): $T^4=2^{13}$ and $H\simeq -31.5$.
(b): $T=2$ and $H\simeq -60$.}
\label{lowtailpath}
\end{figure}

Figure \ref{lowtailpath} a shows 10 sampled configurations at $\tau = t/2$, conditioned on a large negative $H$ (close to -31.5) at
$\tau =t$. They appear in the Gaussian random ensemble with
a probability near $10^{-1000}$.
Here too we only presented the average profile with error bars. Again, the error bars are very small, so that the  tube of stochastic paths around the optimal path is very narrow, clearly indicating the presence of a well-defined optimal path. The profile exhibits an extended flat region, as predicted by Eq.~(\ref{hin1}). The quantitative agreement with the leading-order zero-diffusion  prediction (\ref{hin1}) and (\ref{hout1}) is fairly good \cite{error}.  It should further improve if one
solves Eqs.~(\ref{eqh}) and ~(\ref{eqrho}) for this $H$ numerically, rather than rely on the leading-order asymptotic at $|H|\gg 1$.

As we already mentioned, the OFM correctly predicts the leading-order tails (\ref{hightail}) and (\ref{lowtail}) at arbitrary long times \cite{DMS2016,SMP,Corwinetal,Tsai}. Do the optimal paths still dominate the contribution to the height probability at long times? To answer this question we simulated the directed polymer at much lower temperature, $T=2$, when $\epsilon
=(t/t_{\text{NL}})^{1/2}=10^{5.5}\gg 1$. At this temperature the continuous stochastic heat equation~(\ref{A010}) is not expected to accurately approximate the exact recursive equation~(\ref{A004}).  However, we can still expect an emergence of a well-defined optimal path if we condition the \emph{discrete} polymer on a large deviation of its free energy.  This is indeed what our simulations clearly show, see Fig. \ref{lowtailpath} b.
These configurations appear in the Gaussian random ensemble with
a probability near $10^{-160}$.
Again, the error bars are very small, implying a well-defined optimal path. The quantitative agreement
with Eqs.~(\ref{hin1}) and (\ref{hout1}) is still fair, although $h$ decreases with $|x|$ at large $|x|$ faster than predicted. This latter  effect reflects a  systematic difference in the behaviors of the deterministic solutions of
the discrete equation~(\ref{A004}) and of its continuous approximation (\ref{A010}).

\textit{5. In conclusion}, the problem of one-point height
statistics of the KPZ equation clearly demonstrates the versatility and robustness of the OFM. Importantly, the optimal path can persist, and the OFM can be applicable, in the distribution tails,  well beyond the weak-noise limit. From a broader perspective, our findings extend previous observations of optimal paths in low-dimensional noisy dynamical system out of equilibrium \cite{Dykman1992,Luchinsky,Hales,Ray,Dykman2008} to the realm of non-stationary stochastic fields.

\textit{Acknowledgments}. We thank Alexandre Krajenbrink,
Alberto Rosso and Gregory Schehr
for useful discussions. The simulations were performed in Oldenburg on the HPC cluster CARL which is funded by the DFG
through its Major Research Instrumentation Programme (INST 184/157-1 FUGG) and the Ministry of
Science and Culture (MWK) of the Lower Saxony State. B.M. acknowledges financial support from the Israel Science Foundation (grant No. 807/16). P.S.'s work is supported by the project ``High Field Initiative" (CZ.02.1.01/0.0/0.0/15\_003/0000449) of the European Regional Development Fund.

\end{document}